\documentclass[11pt,a4paper,twoside,graphicx,color]{article}
\usepackage[margin=2cm]{geometry}
\usepackage{graphicx}
\usepackage{txfonts}
\usepackage{natbib}
\begin{document}
 
\begin{center}{\huge \bf
%-------------------------------------------------------------------
%Article Title
Fueling processes on (sub-)kpc scales
%-------------------------------------------------------------------
}\end{center}
\centerline{\bf Francoise Combes}

\smallskip
Observatoire de Paris, LERMA, Coll\`ege de France, CNRS, PSL University, Sorbonne University, 75014, Paris 
e-mail: francoise.combes@obspm.fr

\bigskip
{\bf Abstract:} Since the 1970s, astronomers have struggled with the issue of how matter
can be accreted to promote black hole growth. While low-angular-momentum
stars may be devoured by the black hole, they are not a sustainable source
of fuel. Gas, which could potentially provide an abundant fuel source,
presents another challenge due to its enormous angular momentum.
While viscous torques are not significant, gas is subject to gravity torques
from non-axisymmetric potentials such as bars and spirals. Primary bars
can exchange angular momentum with the gas inside corotation, driving
it inward spiraling until the inner Lindblad resonance is reached.
An embedded nuclear bar can then take over. As the gas reaches the black
hole's sphere of influence, the torque turns negative, fueling the center.
Dynamical friction also accelerates the infall of gas clouds closer to
the nucleus. However, due to the Eddington limit, growing a black hole from
a stellar-mass seed is a slow process. The existence of very massive black
holes in the early universe remains a puzzle that could potentially be
solved through direct collapse of massive clouds into black holes or
super-Eddington accretion.


\section{Introduction}

  One of the main issues to explain the fueling of active galactic nuclei (AGN)
  is the mechanism to get rid of the angular momentum (AM). The black hole may swallow 
  the neighbouring stars, which will create a depletion among those with a low AM. 
  It will require a long relaxation time to replenish this loss cone, so that the
  fueling will rely on the gas infall. The latter requires gravity torques,
  tangential forces, and therefore non axisymmetric features, such as bars or spirals.
  Large-scale features are not sufficient, and should be supported by embedded bars to prolong
  their action towards the center.

  The high spatial resolution provided by the ALMA interferometer has been able
  to reveal these embedded structures, in particular nuclear spirals inside the
  nuclear rings, corresponding to the inner Lindblad resonance (ILR) of the bars.
  Examples of these structures have been unveiled in nearby Seyfert galaxies.

 Inside these nuclear spirals, at 10~pc scale, molecular tori have been unveiled,
 as circumnuclear disks, kinematically decoupled from the large-scale disk. The 
 origin of the decoupling may be found in several causes, like accretion of gas with
 different AM, and/or precession and warping of the very central disk due to relativistic
 effects, coupled with the supermassive black hole spin.

\section{Angular Momentum Problem}

\begin{figure}[hb!]
\includegraphics[clip,width=0.9\textwidth,angle=0]{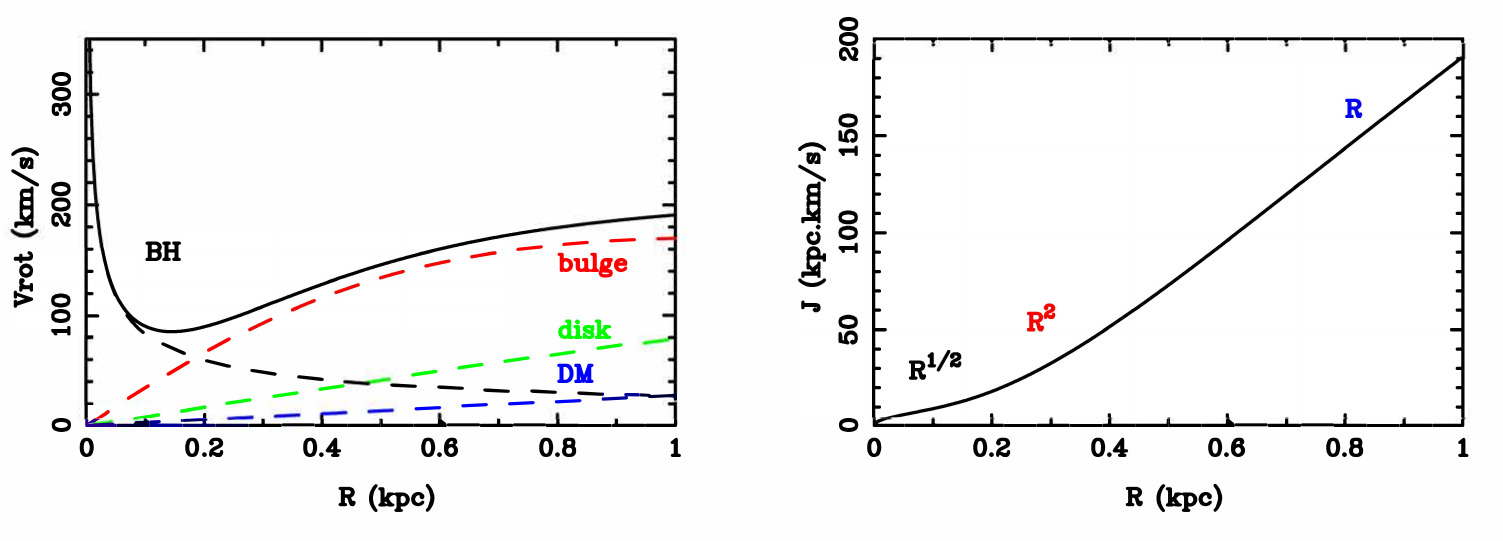}
\caption{How does the angular momentum increase with radius around a super-massive
	black hole (BH)?: {\bf left}, the typical rotation curve in a spiral galaxy is
	dominated by the BH mass in the center (keplerian potential), then is rising due
	to the bulge, and finally ends as flat due to the conspiracy of the disk and 
	dark matter halo (DM). These various contributions are marked in different
	colors; {\bf right},
	the corresponding angular momentum per unit mass increases as power-laws 
	of the radius, with slopes first one-half (Keplerian), than two and one. 
	}
\label{fig1}
\end{figure}

The problem comes from the large contrast between the angular
momentum of the gas in the last stable orbit, 
 L = 2$\times$10$^{24}$ (M/10$^8$M$_\odot$) cm$^2$/s, for a typical black hole
 mass of 10$^8$M$_\odot$, and the gas AM at 3~kpc, L = 10$^{29}$ cm$^2$/s.
The ratio of these two values means that the gas has to lose
5 orders of magnitude in AM, in a relatively short dynamical time. The illustration 
of the AM increase with radius is shown in Fig. \ref{fig1}. For a typical 
luminosity of 10$^{46}$ erg/s, the central
engine has to swallow 2M$_\odot$/yr, during a duty cycle of 100 Myr.

Stars with low AM in the neighborhood of the black hole
can be tidally sheared, and their gas be accreted in a rotating 
disk. These TDE (Tidal Disruption Event) may occur at the frequency of 
one every 10 000 yrs in the Milky Way. Some have been detected in
external galaxies, their signature being a characteristic light curve
decreasing with time as a -5/3 power-law \citep{Miller2015}.
But soon, the depletion and loss cone effect dries up this fueling source,
unless galaxy interactions re-shuffled the stellar distribution, 
and creates nuclear star clusters, in a nuclear starburst.

Gas is however the main fuel, and is driven inwards
through non-axisymmetries of the galaxy potential.
 Several steps can be distinguished in this process.
First, primary bars of typical diameters 10~kpc, can
drive the gas from their corotation to R $\sim$ 100~pc, where
the gas may be stalled in a ring. 
Then embedded nuclear bars can prolong their action
from 100~pc to 10~pc. Non-axisymmetries are 
mainly of m=2 morphology, but also m=1 (lopsidedness), 
or tidal forces from companions can play a role.

At smaller scales of ~1-10~pc, other processes also have to be
considered: turbulence, viscosity, warps, bends,
  dynamical friction, formation of thick disks, as long as gas remains
  in sufficient amount.

\begin{figure}[ht!]
\includegraphics[clip,width=0.9\textwidth,angle=0]{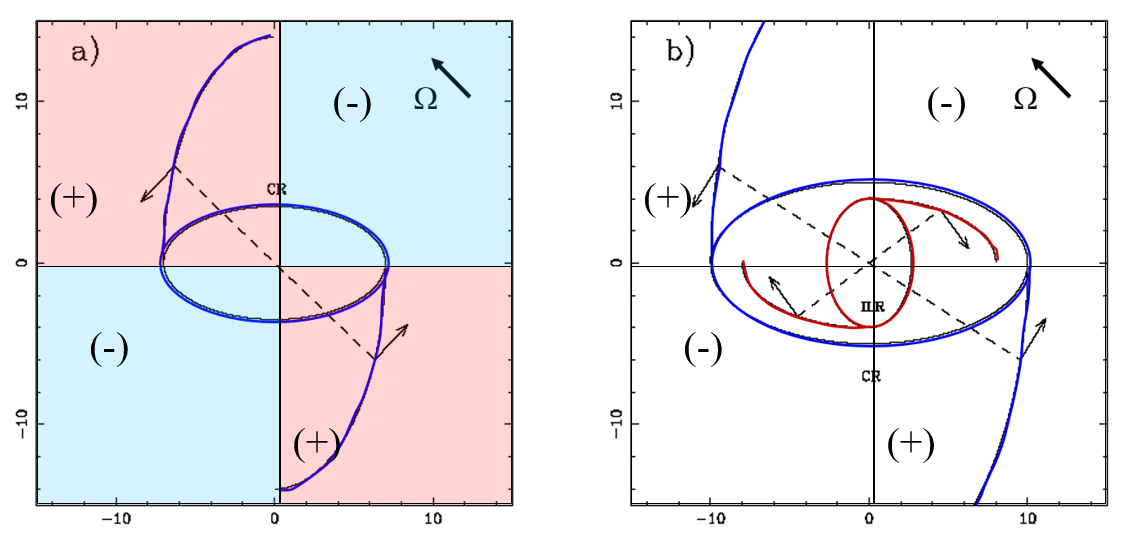}
\caption{The sign of the torques exerted by the bar on the gas can be obtained geometrically
	through these schematic diagrams: {\bf left},  outside corotation, or roughly
	outside the bar, the torque is positive (red quadrant) with respect to the
	sense of rotation ($\Omega$), and the gas is driven out toward the outer Lindblad 
	resonance (OLR);  {\bf right},  inside the corotation, the torque is negative 
	(blue quadrant), and the gas is driven inward, down to the inner resonance (ILR).
	}
\label{fig2}
\end{figure}

\subsection{Dynamics of bars}

 Let us recall the main features of barred galaxies:
the stellar orbits are classified through a skeleton of
periodic orbits. The latter 
are orbits that close on themselves after one or more turns in the bar 
rotating frame. They are the building blocks which determine the stellar 
distribution function, since they define families of trapped orbits around them. 
Trapped orbits are non-periodic, but oscillate about one periodic orbit, 
with a similar shape. 
The periodic orbits are numerous (see the review by \citet{Contopoulos1989}), 
let us mention here the most important ones for the bar support. Inside corotation, 
the x1 family is the main family supporting the bar. Orbits are elongated parallel 
to the bar, within corotation. There exists also the x2 family, but only between 
the two inner Lindblad resonances (ILR), if they exist. They are more round, and 
elongated perpendicular to the bar. The existence of two ILR's in the axisymmetric 
sense might not be sufficient for the x2 family to appear. When the bar is 
strong enough, the x2 orbits disappear. The bar strength necessary to eliminate 
the x2 family depends on the pattern speed, the lower this speed, 
the stronger the bar must be.
Outside corotation, the 2 / 1 orbits that are run in the retrograde sense in the 
rotating frame are perpendicular to the bar inside the outer Lindblad resonance (OLR), 
and parallel to the bar slightly outside (see discussion by \citet{Kalnajs1991}). 
Their shape reveals a characteristic figure-eight, that is very similar to the 
dimpled shape of some outer rings in barred galaxies. 

In summary, periodic orbits are aligned parallel or
perpendicular to the bar, and their orientation changes
by 90$^\circ$ at each resonance
\citep{Contopoulos1980, Athanassoula1992}.
Gas tends to follow periodic orbits, but its
dissipative character, due to cloud collisions,
means that orbits cannot cross. Instead their
orbits are tilted, and they change gradually by 
 90$^\circ$ at each resonance. The crowding of these
 stream lines produces a spiral morphology
\citep{Sanders1976}. The spiral is open, and at
maximum can rotate by 180-360$^\circ$.

\begin{figure}[hb!]
\includegraphics[clip,width=0.9\textwidth,angle=0]{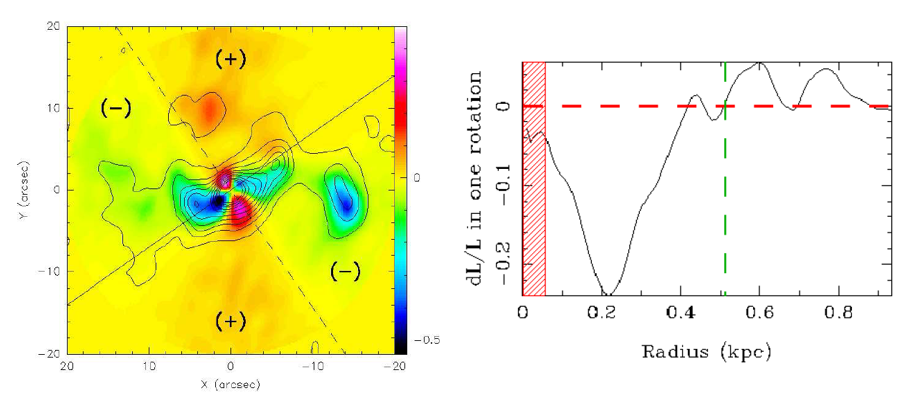}
\caption{The gravity torques can be measured on real galaxies, by computing in each
	pixel the potential and forces from a red image (old stars), and comparing with the 
	observed gas density. The {\bf left} image is the torque map in color, with the bar
	splitting the plane in four quadrants, for NGC 3627. The gas density is overlaid in
	contours. The {\bf right} plot quantifies the relative angular momentum lost in one 
	rotation, while averaging the torque over azimuth, weighted by gas density. Adapted 
	from \citet{Casasola2011}.
	}
\label{fig3}
\end{figure}   

\subsection{Embedded structures}

Because the gas and the stars are not in phase, stellar
bars exert a torque on the gas, except at resonances,
where the gas piles up and stalls in rings, aligned with the bar
in some way. When mass has been accumulating in the center,
all frequencies $\Omega$ and $\Omega-\kappa/2$, the orbit precessing rate,
see their value significantly increased, implying the existence of two ILR. In between
these, the periodic orbits (x2) are perpendicular to the bar. Stars cannot
sustain the bar anymore, and weaken it. The z-resonance, creating peanut-shape
bulges, also weakens the bar.
This triggers the decoupling of a second bar, an embedded nuclear
bar \citep{Friedli1993}. This second bar rotates faster than the primary
bar, and both are misaligned \citep{Buta1996}.

\begin{figure}[ht!]
\includegraphics[clip,width=0.9\textwidth,angle=0]{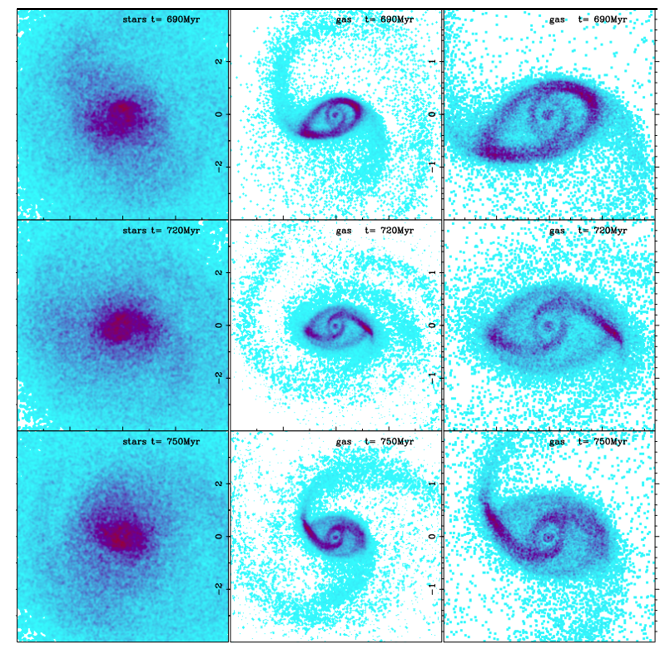}
	\caption{Hydro-N-body simulation of a double bar in a spiral galaxy. The {\bf left}
	panel shows the stellar component surface density (linear scale, axes in kpc),
    the {\bf middle} panel shows the gas component (log scale), while the {\bf right} panel
    is also the gas, in a zoomed spatial scale. The simulation shows clearly that 
	the gas first piles up in the external ring (ILR of the primary bar), then 
	progressively infalls inside the ILR to gather at the nuclear bar resonance.
	Adapted from \citet{Hunt2008}.
	}
\label{fig4}
\end{figure}   

\subsection{Fueling AGN: removing angular momentum (AM)}

The torques exerted on the gas by the bar change sign at
each resonance, cf Fig. \ref{fig2}.  Outside the corotation radius (CR), 
the gas is driven outwards and accumulates at the outer Lindblad 
resonance (OLR). Inside CR, the gas is driven inwards, at least down to ILR.
To quantify the phenomenon, bar gravity torques can be 
computed from the red images tracing old stars, and the potential.
When the gas distribution is overlaid on the torque map (cf Fig. \ref{fig3}),
the sign and amount of the gas infall can be estimated, as in
NGC3627 \citep{Casasola2011}. The correlation
 between bars and AGN is still debated 
\citep{Schawinski2010,Masters2011,Cardamone2011,Alonso2014}, 
since the time-scales corresponding
to the primary and secondary bars are quite different, and also
different from the AGN duty cycle.

During a survey of about 20 galaxies with the IRAM
interferometer, statistics of fueling at the 
 10-100~pc scale was obtained \citep{Garcia-Burillo2012}.
Only ~35\% of negative torques were measured in the center, 
the rest of the times, positive torques were measured, 
or gas was stalled in a ring. In this case, future fueling
has to wait for the decoupling of a secondary bar, as shown 
in the simulation of Fig. \ref{fig4}. This means that
the fueling phases are short, a few 10$^7$ yrs, and may be due to feedback.
Star formation is also fueled by the torques, and is
always associated with AGN activity, but with longer time-scales.

\section{Small-scale fueling with ALMA}

\begin{figure}[hb!]
\begin{center}
\includegraphics[clip,width=0.7\textwidth,angle=0]{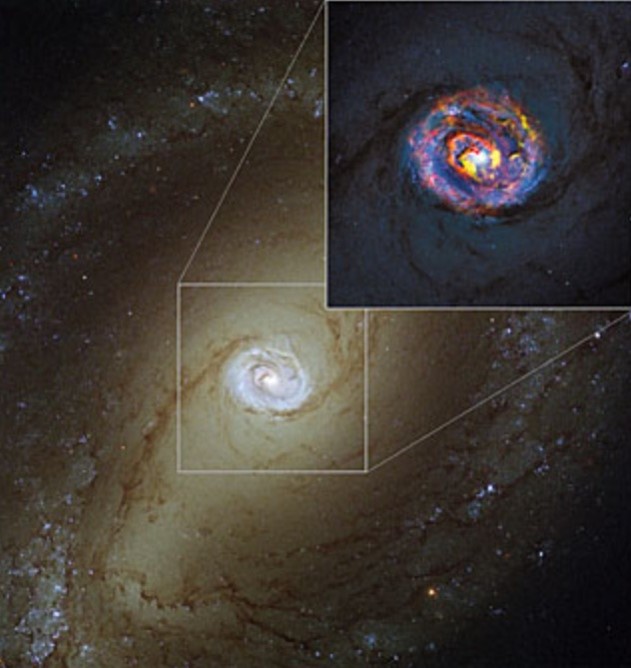}
\end{center}
\caption{HST image of NGC1433 bar, with characteristic leading dust lanes,
	ending down to a blue ring at the inner Lindblad resonance (ILR). The insert 
	displays the blue ring from HST overlaid with the orange-red image from ALMA
	of the molecular gas, traced by its CO(3-2) emission. The latter reveals a
	second ring inside the first ILR. Adapted from \citet{Combes2013}.
	}
\label{fig5}
\end{figure}

With the advent of ALMA, higher spatial resolution is possible,
to explore the 10~pc scales in nearby galaxies. One of the first barred 
spiral observed was NGC 1433 (cf Fig. \ref{fig5}). While only a star forming
ring was observed at ILR with HST, a second ring was detected in the molecular gas 
with ALMA, corresponding to the second ILR. The computation of the torques
have shown that the AGN is not presently fueled, but positive torques bring the
gas from the center to the second ring. Negative torques outside this second ring
contribute to accumulate gas there \citep{Smajic2014}.
On the minor axis, an outflow has been detected, with small velocity.
It is one of the smallest outflows detected, with M(H$_2$)= 3.6$\times$10$^6$ M$_\odot$ and
a flow rate of 7 M$_\odot$/yr \citep{Combes2013}.

\begin{figure}[ht!]
\begin{center}
\includegraphics[clip,width=0.7\textwidth,angle=0]{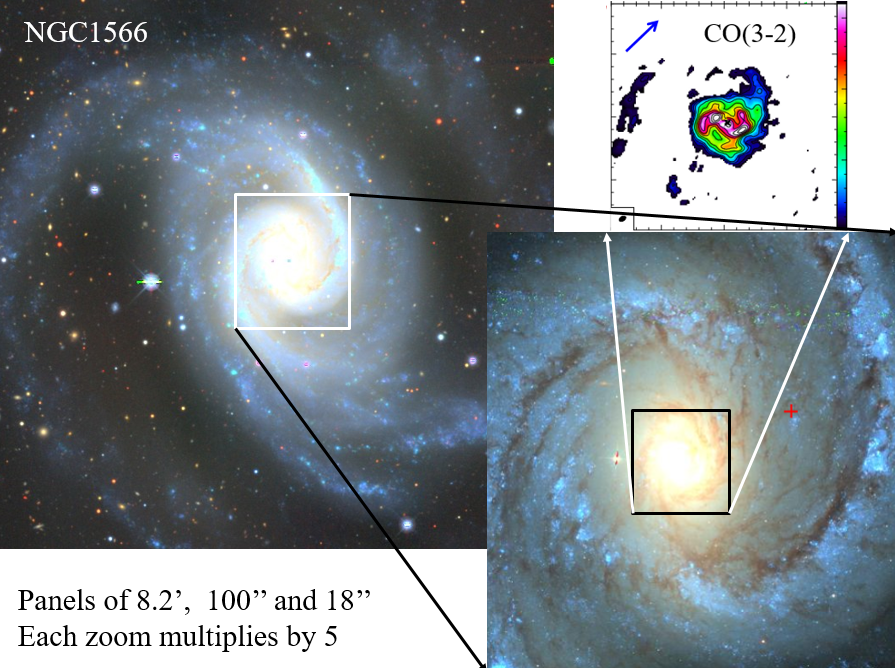}
\end{center}
\caption{The barred spiral galaxy NGC 1566 reveals several embedded structures, as
        seen in these three progressively zoomed images. The last one reveals a
	trailing nuclear spiral in the molecular gas, traced by its CO(3-2) emission
	observed with ALMA. Adapted from \citet{Combes2014}.
	}
\label{fig6}
\end{figure}

\begin{figure}[hb!]
\includegraphics[clip,width=0.9\textwidth,angle=0]{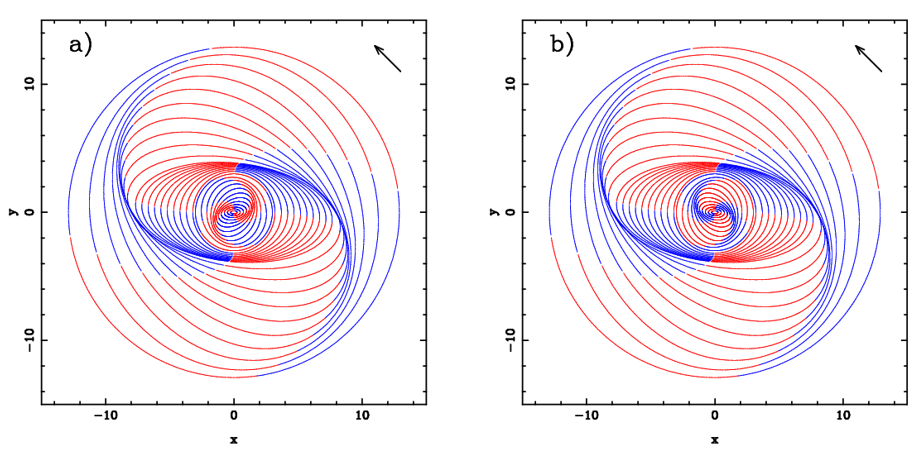}
\caption{Gas streamlines schematically represented by elliptical orbits, precessed
	to form logarithmic spirals. They are initially aligned on the horizontal
	axis (parallel to the bar) and colored according to the four quadrants. The pattern
	speed is such that there exists an ILR inside the bar, delineated by a ring. 
	{\bf Left:} without a super-massive black hole (BH), the precessing rate 
	decreases towards the center, and the gas forms a leading nuclear spiral.
	{\bf Right:} with a BH dominating the potential inside the ILR, the precessing rate 
	increases towards the center, and the gas forms a trailing spiral. This changes the
	sign of the torque exerted on the gas.
	}
\label{fig7}
\end{figure}

Other barred Seyfert galaxies, like NGC1566 were found
in the feeding phase (cf Fig. \ref{fig6}). Inside the ILR ring,
the CO(3-2) emission map from ALMA reveals the existence of 
a nuclear disk, with a trailing nuclear bar. The existence 
of the trailing spiral was a surprise, since without
the gravitational influence of the central black hole, a leading 
spiral is expected. This is due to the shape of the $\Omega-\kappa/2$
curve as a funciton of radius. When the gas infalls, it precesses more
rapidly if the curve is climbing when radius is reduced, which means
a trailing arm, and the contrary if the curve declines, cf Fig. \ref{fig7}.
 If the spiral is trailing, it means that the gas feels a climbing curve,
 due to the BH, and it has entered the sphere of influence of the BH.
The computation of the torques in
NGC1566 has confirmed the evidence of fueling \citep{Combes2014}.

In several other barred galaxies, a trailing nuclear spiral has
been revealed in the molecular component with ALMA, for instance
in NGC 1808 or NGC 613. This indicates that the gas has entered
the sphere of influence of the black hole, making the torque 
negative, ensuring the fueling of the AGN.
This trailing spiral develops always inside the ILR ring 
of the bar \citep{Audibert2021}.

NGC 613 is the academic case of a strong
barred galaxy, with a star forming gas ring at ILR. 
The first ALMA observations, with moderate spatial resolution,
could see only the ring, without resolving the internal 
structure \citep{Miyamoto2017}. With a beam of 60mas, it
was possible to clearly see an internal trailing nuclear
spiral, and inside the spiral a molecular torus (cf Fig. \ref{fig8}).
The computation of the torques indicates that the gas can lose all 
its angular momentum in only one rotation, when inside 50~pc radius
\citep{Audibert2019}. In addition, a molecular outflow is
detected along the minor axis, parallel to the cm-wave detected radio jet.

\section{Molecular tori}

Our vision of the central regions of AGN
surrounding the black hole have changed significantly
in the recent years. For a long time, around the accretion disk and the
broad line region (BLR), a dusty torus, with a donut shape was
assumed to exist. It was thought to obscure the BLR for observers
seeing the accretion disk almost edge-on. But infrared interferometers have
shown that the dust is frequently detected in the polar direction, instead
of being aligned along the putative torus. This polar dust must be ejected
through the AGN wind, starting from the dust sublimation radius (pc size),
and forming the border of a hollow cone \citep{Hoenig2019}.
 The gas motion is then both infall in a thin disk, forming a molecular
 torus, and then outflow in the perpendicular direction.

\begin{figure}[ht!]
\includegraphics[clip,width=0.9\textwidth,angle=0]{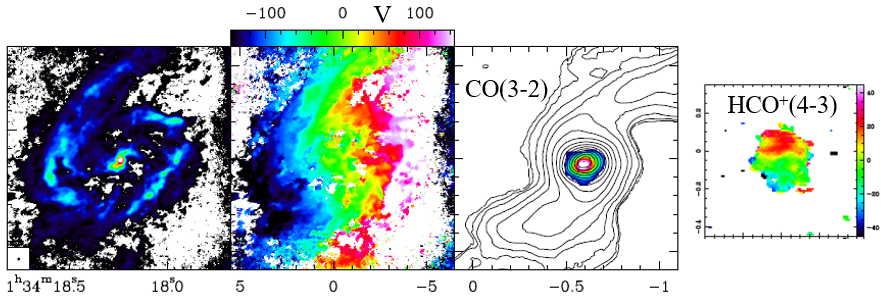}
\caption{The barred galaxy NGC 613 shows a contrasted ring at its ILR, in
      the molecular gas traced by CO(3-2) with ALMA. From left to right is
	the gas surface density, then the velocity field, and a 10-fold zoom
	of the gas density (contours), with the radio continuum (color-scale) at the
	center. The right-most panel is the velocity field of HCO$^+$(4-3) emission,
	revealing a misalignment with the large-scale gas.
	Adapted from \citet{Combes2019} and \citet{Audibert2019}.
	}
\label{fig8}
\end{figure}

Figure \ref{fig8} shows that the molecular torus, inserted in the nuclear
spiral, has a decoupled kinematics, i.e. the kinematic major axis is
not aligned with that of the large-scale disk \citep{Combes2019,Audibert2019}.

\begin{figure}[hb!]
\includegraphics[clip,width=0.9\textwidth,angle=0]{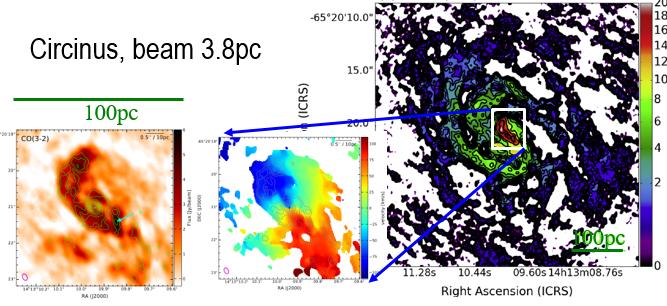}
	\caption{Image of the molecular gas, obtained in CO(3-2) with ALMA, of
	the Circinus galaxy, with a resolution of 3.8~pc. There is a nuclear spiral, 
	of 20-30~pc, and inside another circum-nuclear disk, which could be the 
	molecular torus  (edge-on, for this type 2 AGN).
	Adapted from \citet{Tristram2022}.
	}
\label{fig9}
\end{figure}

This misalignment is frequent in all nearby galaxies
observed with ALMA with 10~pc resolution, such as
NGC 1672 or NGC 1326 in the NUGA sample \citep{Combes2019},
or NGC 5643, NGC 6300 in the GATOS sample 
\citep{Garcia-Burillo2021}. It is yet not possible to
determine whether the central molecular torus is simply
continuously warped, or tilted, with discontinuous
disks torn into a few pieces. 
Circinus is one of the nearest galaxies, observed down to 2~pc resolution,
and shows a nuclear spiral, of 20-30~pc, with inside another circum-nuclear
disk, which can be interpreted as the molecular torus. It is self-absorbed
in CO(3-2) but not in CO(6-5), nor in the dense gas tracers, such as HCO$^+$.
In addition to this cold thin disk, fueling the AGN, there is an outflow in the polar
direction, composed of warm dust and ionized gas, cf Fig. \ref{fig9}
\citep{Izumi2018, Tristram2022}.

The prototypical Seyfert 2 NGC 1068 has been intensively observed,
with high resolution at many-wavelengths. While the near-infrared
reveals clearly the hollow polar cone in warm dust \citep{Gratadour2015},
the cold component in the millimeter domain reveals a
molecular torus of 7-10~pc in diameter
\citep{Garcia-Burillo2016, Imanishi2020}.
The torus is quasi edge-on, misaligned with the large-scale
disk, warped, more inclined than the water maser disk,
and maybe suffers from counter-rotation
\citep{Impellizzeri2019, Imanishi2020}.
In contrast, the barred Seyfert 1 galaxy NGC 1097, with ALMA
at ~10~pc resolution, does not show any dense and compact torus.
There is a nuclear spiral inside the star-forming ring at the ILR 
of the bar \citep{Fathi2006, Izumi2017}.
The absence of torus concerns about 10-20\% of low-luminosity
AGN. This percentage increases with the Eddington ratio,
and could be related to the AGN feedback
\citep{Garcia-Burillo2021}.

\begin{figure}[ht!]
\includegraphics[clip,width=0.9\textwidth,angle=0]{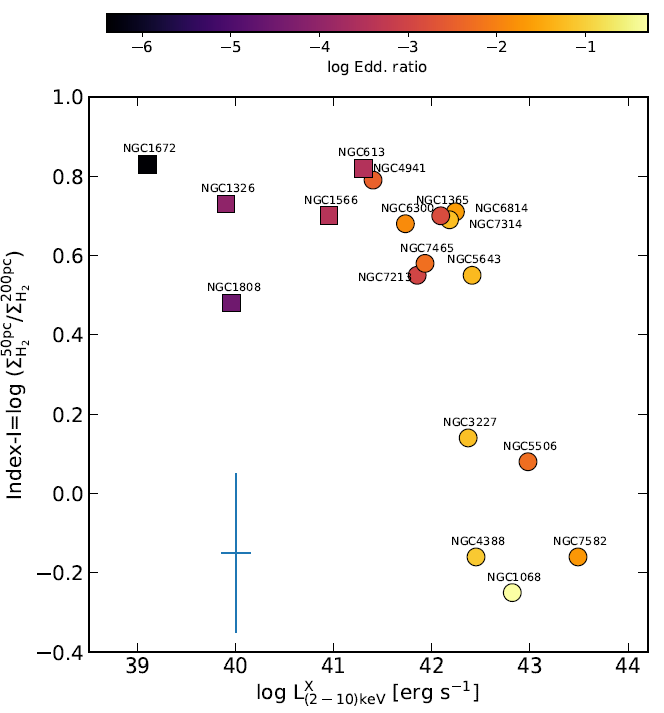}
\caption{Concentration of molecular gas in the central regions of galaxies, versus
 their AGN luminosities measured in the 2–10 keV X-ray band. The concentration is
	measured from the ratio of the average H$_2$ surface densities at two spatial scales: 
	r$<$50~pc and r$<$ 200~pc, characteristic of the nuclear and circumnuclear regions. 
        Symbols are color-coded as a function of the Eddington ratios.
	The sample galaxies: NUGA (square markers) and GATOS (circle markers) can
	be separated both by their AGN activity, and molecular gas concentration.
	Adapted from \citet{Garcia-Burillo2021}.
	}
\label{fig10}
\end{figure}

The influence of the AGN activity is indeed visible
on the molecular gas concentration. The latter can be
quantified by the surface density ratio of the molecular
gas inside 50~pc and inside 200~pc. The most active AGN 
have less H$_2$ concentration. The latter drops for
X-ray luminosities Lx$>$ 10$^{42.5}$ erg/s, or for
Eddington ratios $>$ 10$^{-3}$, see Fig. \ref{fig10},
\citep{Garcia-Burillo2021}.

\section{Early black holes}

Along their life-time, galaxies grow both their bulge and their black hole,
in synergy, so that a tight relation can be observed between both masses
\citep{Kormendy2013}. Secular evolution, together with some tidal interaction with
companions, can provide the fuel for AGN and BH growth. However, AGN feedback
makes the BH growth intermittent, as described above.
Some different processes should occur at high redshift, to account
for the observations of high black hole masses (M$>$ 10$^9$ M$_\odot$) already at z$>$6
\citep{Venemans2016}.
 Hundreds of quasars have been detected at high redshift, z $>$6, with black 
hole masses between 10$^8$ and 10$^{10}$ M$\odot$, in the first billion years 
of the Universe \citep[e.g.][]{Mortlock2011,Yang2020, Farina2022}.
The black hole growth rate through accretion is proportional to its mass, and is thus very slow
at the beginning, starting with stellar mass seeds. 
At any time, the Eddington luminosity
imposes an upper limit in the accretion rate, limit which is 
proportional to the BH mass. 
In contrast, at high redshift, black holes appear to grow faster than their 
host bulges, and the proportionality relation breaks.
Several solutions have been proposed. One is to assume that the BH in the early times 
accrete exceptionally faster than their Eddington rate, and this without being 
stopped by AGN feedback.
 Simulations of black hole growth with super-Eddington accretion, even in a 
massive over-density environment, progenitor of a cluster, in general do not 
succeed to reproduce the observed massive black holes at z=6 \citep{Sassano2023}, 
nor explain why the black hole growth is more rapid than the stellar mass growth, 
contrary to what is observed at z=0. However, with some modifications of the usual 
scenario, taking into account less efficient AGN feedback, more efficient accretion, 
and starting earlier with a more massive seed, it is possible to account for the 
observations \citep{Bennett2024}.

An alternative solution is to assume that the BH collapses directly from a massive cloud,
due to primordial metal abundance, and lack of fragmentation. The seeds to form the 
first black holes would then be much more
massive than stellar, and the growth rate would be strong since the start. This however assumes
contrived suppression of H$_2$ molecule formation, and remains debated. 

The existence of
intermediate mass black hole (IMBH) is part of the puzzle, since they have not yet
been detected unambiguously. The direct collapse scenario could explain their low
abundance. 
The major challenge for observations of IMBHs is their small mass and small 
impact in their surroundings.  Since the radius of influence of a black hole scales 
linearly with its mass, it requires very high spatial resolution, to detect their 
impact on the gas or stellar component. Also the dynamical friction time-scale for 
a black hole to decay towards nuclei is inversely proportional to the mass, and 
IMBH might not have time to infall within a Hubble time \citep[e.g.][]{diMatteo2023}. 

An promising way to detect IMBH is to search for AGN signatures, and candidates
have been observed in several low-mass galaxies \citep{Chilingarian2018}.
There exist candidates for IMBH in several dwarf active galaxy nuclei 
\citep[e.g.][]{Mezcua2017}, however in the core of globular clusters, 
where they were expected, only a collection of stellar-mass black holes 
have been detected instead \citep{Vitral2021,Vitral2023}.

\section{Summary}

The AGN fueling requires non-axisymmetries in the galaxy
potential to drive the gas inwards. The primary bar in
spiral galaxies can drive the gas from its corotation to the inner
Linblad resonance, where the gas is accumulating, and forms a 
starbursting ring, at the scale of 100~pc. Further
fueling has to await the formation of a nuclear bar, embedded
inside the ILR ring. ALMA has revealed 10~pc scale structures
inside the ring, in particular trailing nuclear spirals,
indicating that the gas has entered the sphere of influence of 
the central black hole. The stellar potential then prolongs the
action of the primary bar to fuel the AGN. Inside the nuclear
spiral a molecular disk, kinematically decoupled,
iis often found, that can be interpreted as the molecular torus. 

The misalignment of the torus can be explained by gas accretion
with different angular momentum. This occurs naturally
through star formation/supernovae feedback, which ejects
gas above the plane. When this gas comes back, by the fountain 
effect, it can arrive in any direction with a random angular
momentum. Numerical simulations have shown that even
polar gas rings or disks can be formed
\citep{Renaud2015, Emsellem2015}.

The decoupling of molecular tori is not too surprising,
given the very different dynamical time-scales between the 10~pc 
and 100~pc scales. In addition, the material at much less than 1~pc
from the black hole is certainly influenced by relativistic effects, related
to the black hole spin: the Bardeen-Petterson effect, where the torque
exerted by the BH tends to align the material perpendicular
to its spin, and produces warps. Warping might also be
radiation-driven \citep{Pringle1996,Maloney1997}.

 Contrary to the existence of a tight relation between bulge and black 
hole masses in nearby galaxies, super-massive black holes grow relatively 
faster than the stellar component in early galaxies, in the first billion 
year of the Universe. To reproduce this behavior, simulations have assumed 
super-Eddington accretion, and low-efficiency AGN feedback at these epochs. 
Alternatively, direct collapse of super-massive stars into black holes 
might contribute to the solution. Future observations in the early universe 
with JWST might enlight the issue.

\bigskip
{\bf Acknowledgements:} Thanks to Ilaria Ruffa and the organisers of the conference
"AGN on the beach" in Tropea, Italy, where this material was discussed.
FC has benefited from the support of the Programme National Cosmologie et Galaxies.
Figures 3-6 and 8-10 are reproduced with permission from 
Astronomy \& Astrophysics, © ESO.

%=====================================
% References, external bibliography
%=====================================
\bibliographystyle{aa}
\bibliography{combesf.bib}

\begin{thebibliography}{48}
\expandafter\ifx\csname natexlab\endcsname\relax\def\natexlab#1{#1}\fi

\bibitem[{{Alonso} {et~al.}(2014){Alonso}, {Coldwell}, \&
  {Lambas}}]{Alonso2014}
{Alonso}, S., {Coldwell}, G., \& {Lambas}, D.~G. 2014, \aap, 572, A86

\bibitem[{{Athanassoula}(1992)}]{Athanassoula1992}
{Athanassoula}, E. 1992, \mnras, 259, 328

\bibitem[{{Audibert} {et~al.}(2019){Audibert}, {Combes}, {Garc{\'\i}a-Burillo},
  {Hunt}, {Eckart}, {Aalto}, {Casasola}, {Boone}, {Krips}, {Viti}, {Muller},
  {Dasyra}, {van der Werf}, \& {Mart{\'\i}n}}]{Audibert2019}
{Audibert}, A., {Combes}, F., {Garc{\'\i}a-Burillo}, S., {et~al.} 2019, \aap,
  632, A33

\bibitem[{{Audibert} {et~al.}(2021){Audibert}, {Combes}, {Garc{\'\i}a-Burillo},
  {Hunt}, {Eckart}, {Aalto}, {Casasola}, {Boone}, {Krips}, {Viti}, {Muller},
  {Dasyra}, {van der Werf}, \& {Mart{\'\i}n}}]{Audibert2021}
{Audibert}, A., {Combes}, F., {Garc{\'\i}a-Burillo}, S., {et~al.} 2021, \aap,
  656, A60

\bibitem[{{Bennett} {et~al.}(2024){Bennett}, {Sijacki}, {Costa}, {Laporte}, \&
  {Witten}}]{Bennett2024}
{Bennett}, J.~S., {Sijacki}, D., {Costa}, T., {Laporte}, N., \& {Witten}, C.
  2024, \mnras, 527, 1033

\bibitem[{{Buta} \& {Combes}(1996)}]{Buta1996}
{Buta}, R. \& {Combes}, F. 1996, \fcp, 17, 95

\bibitem[{{Cardamone} {et~al.}(2011){Cardamone}, {Schawinski}, {Masters},
  {Lintott}, \& {Fortson}}]{Cardamone2011}
{Cardamone}, C.~N., {Schawinski}, K., {Masters}, K., {Lintott}, C., \&
  {Fortson}, L. 2011, in American Astronomical Society Meeting Abstracts, Vol.
  218, American Astronomical Society Meeting Abstracts \#218, 206.03

\bibitem[{{Casasola} {et~al.}(2011){Casasola}, {Hunt}, {Combes},
  {Garc{\'\i}a-Burillo}, \& {Neri}}]{Casasola2011}
{Casasola}, V., {Hunt}, L.~K., {Combes}, F., {Garc{\'\i}a-Burillo}, S., \&
  {Neri}, R. 2011, \aap, 527, A92

\bibitem[{{Chilingarian} {et~al.}(2018){Chilingarian}, {Katkov}, {Zolotukhin},
  {Grishin}, {Beletsky}, {Boutsia}, \& {Osip}}]{Chilingarian2018}
{Chilingarian}, I.~V., {Katkov}, I.~Y., {Zolotukhin}, I.~Y., {et~al.} 2018,
  \apj, 863, 1

\bibitem[{{Combes} {et~al.}(2019){Combes}, {Garc{\'\i}a-Burillo}, {Audibert},
  {Hunt}, {Eckart}, {Aalto}, {Casasola}, {Boone}, {Krips}, {Viti}, {Sakamoto},
  {Muller}, {Dasyra}, {van der Werf}, \& {Martin}}]{Combes2019}
{Combes}, F., {Garc{\'\i}a-Burillo}, S., {Audibert}, A., {et~al.} 2019, \aap,
  623, A79

\bibitem[{{Combes} {et~al.}(2013){Combes}, {Garc{\'\i}a-Burillo}, {Casasola},
  {Hunt}, {Krips}, {Baker}, {Boone}, {Eckart}, {Marquez}, {Neri}, {Schinnerer},
  \& {Tacconi}}]{Combes2013}
{Combes}, F., {Garc{\'\i}a-Burillo}, S., {Casasola}, V., {et~al.} 2013, \aap,
  558, A124

\bibitem[{{Combes} {et~al.}(2014){Combes}, {Garc{\'\i}a-Burillo}, {Casasola},
  {Hunt}, {Krips}, {Baker}, {Boone}, {Eckart}, {Marquez}, {Neri}, {Schinnerer},
  \& {Tacconi}}]{Combes2014}
{Combes}, F., {Garc{\'\i}a-Burillo}, S., {Casasola}, V., {et~al.} 2014, \aap,
  565, A97

\bibitem[{{Contopoulos} \& {Grosbol}(1989)}]{Contopoulos1989}
{Contopoulos}, G. \& {Grosbol}, P. 1989, \aapr, 1, 261

\bibitem[{{Contopoulos} \& {Papayannopoulos}(1980)}]{Contopoulos1980}
{Contopoulos}, G. \& {Papayannopoulos}, T. 1980, \aap, 92, 33

\bibitem[{{Di Matteo} {et~al.}(2023){Di Matteo}, {Ni}, {Chen}, {Croft}, {Bird},
  {Pacucci}, {Ricarte}, \& {Tremmel}}]{diMatteo2023}
{Di Matteo}, T., {Ni}, Y., {Chen}, N., {et~al.} 2023, \mnras, 525, 1479

\bibitem[{{Emsellem} {et~al.}(2015){Emsellem}, {Renaud}, {Bournaud},
  {Elmegreen}, {Combes}, \& {Gabor}}]{Emsellem2015}
{Emsellem}, E., {Renaud}, F., {Bournaud}, F., {et~al.} 2015, \mnras, 446, 2468

\bibitem[{{Farina} {et~al.}(2022){Farina}, {Schindler}, {Walter},
  {Ba{\~n}ados}, {Davies}, {Decarli}, {Eilers}, {Fan}, {Hennawi},
  {Mazzucchelli}, {Meyer}, {Trakhtenbrot}, {Volonteri}, {Wang}, {Worseck},
  {Yang}, {Gutcke}, {Venemans}, {Bosman}, {Costa}, {De Rosa}, {Drake}, \&
  {Onoue}}]{Farina2022}
{Farina}, E.~P., {Schindler}, J.-T., {Walter}, F., {et~al.} 2022, \apj, 941,
  106

\bibitem[{{Fathi} {et~al.}(2006){Fathi}, {Storchi-Bergmann}, {Riffel}, {Winge},
  {Axon}, {Robinson}, {Capetti}, \& {Marconi}}]{Fathi2006}
{Fathi}, K., {Storchi-Bergmann}, T., {Riffel}, R.~A., {et~al.} 2006, \apjl,
  641, L25

\bibitem[{{Friedli} \& {Martinet}(1993)}]{Friedli1993}
{Friedli}, D. \& {Martinet}, L. 1993, \aap, 277, 27

\bibitem[{{Garc{\'\i}a-Burillo} {et~al.}(2021){Garc{\'\i}a-Burillo},
  {Alonso-Herrero}, {Ramos Almeida}, {Gonz{\'a}lez-Mart{\'\i}n}, {Combes},
  {Usero}, {H{\"o}nig}, {Querejeta}, {Hicks}, {Hunt}, {Rosario}, {Davies},
  {Boorman}, {Bunker}, {Burtscher}, {Colina}, {D{\'\i}az-Santos}, {Gandhi},
  {Garc{\'\i}a-Bernete}, {Garc{\'\i}a-Lorenzo}, {Ichikawa}, {Imanishi},
  {Izumi}, {Labiano}, {Levenson}, {L{\'o}pez-Rodr{\'\i}guez}, {Packham},
  {Pereira-Santaella}, {Ricci}, {Rigopoulou}, {Rouan}, {Shimizu}, {Stalevski},
  {Wada}, \& {Williamson}}]{Garcia-Burillo2021}
{Garc{\'\i}a-Burillo}, S., {Alonso-Herrero}, A., {Ramos Almeida}, C., {et~al.}
  2021, \aap, 652, A98

\bibitem[{{Garc{\'\i}a-Burillo} \& {Combes}(2012)}]{Garcia-Burillo2012}
{Garc{\'\i}a-Burillo}, S. \& {Combes}, F. 2012, in Journal of Physics
  Conference Series, Vol. 372, Journal of Physics Conference Series, 012050

\bibitem[{{Garc{\'\i}a-Burillo} {et~al.}(2016){Garc{\'\i}a-Burillo}, {Combes},
  {Ramos Almeida}, {Usero}, {Krips}, {Alonso-Herrero}, {Aalto}, {Casasola},
  {Hunt}, {Mart{\'\i}n}, {Viti}, {Colina}, {Costagliola}, {Eckart}, {Fuente},
  {Henkel}, {M{\'a}rquez}, {Neri}, {Schinnerer}, {Tacconi}, \& {van der
  Werf}}]{Garcia-Burillo2016}
{Garc{\'\i}a-Burillo}, S., {Combes}, F., {Ramos Almeida}, C., {et~al.} 2016,
  \apjl, 823, L12

\bibitem[{{Gratadour} {et~al.}(2015){Gratadour}, {Rouan}, {Grosset},
  {Boccaletti}, \& {Cl{\'e}net}}]{Gratadour2015}
{Gratadour}, D., {Rouan}, D., {Grosset}, L., {Boccaletti}, A., \& {Cl{\'e}net},
  Y. 2015, \aap, 581, L8

\bibitem[{{H{\"o}nig}(2019)}]{Hoenig2019}
{H{\"o}nig}, S.~F. 2019, \apj, 884, 171

\bibitem[{{Hunt} {et~al.}(2008){Hunt}, {Combes}, {Garc{\'\i}a-Burillo},
  {Schinnerer}, {Krips}, {Baker}, {Boone}, {Eckart}, {L{\'e}on}, {Neri}, \&
  {Tacconi}}]{Hunt2008}
{Hunt}, L.~K., {Combes}, F., {Garc{\'\i}a-Burillo}, S., {et~al.} 2008, \aap,
  482, 133

\bibitem[{{Imanishi} {et~al.}(2020){Imanishi}, {Nguyen}, {Wada}, {Hagiwara},
  {Iguchi}, {Izumi}, {Kawakatu}, {Nakanishi}, \& {Onishi}}]{Imanishi2020}
{Imanishi}, M., {Nguyen}, D.~D., {Wada}, K., {et~al.} 2020, \apj, 902, 99

\bibitem[{{Impellizzeri} {et~al.}(2019){Impellizzeri}, {Gallimore}, {Baum},
  {Elitzur}, {Davies}, {Lutz}, {Maiolino}, {Marconi}, {Nikutta}, {O'Dea}, \&
  {Sani}}]{Impellizzeri2019}
{Impellizzeri}, C.~M.~V., {Gallimore}, J.~F., {Baum}, S.~A., {et~al.} 2019,
  \apjl, 884, L28

\bibitem[{{Izumi} {et~al.}(2017){Izumi}, {Kohno}, {Fathi}, {Hatziminaoglou},
  {Davies}, {Mart{\'\i}n}, {Matsushita}, {Schinnerer}, {Espada}, {Aalto},
  {Onishi}, {Turner}, {Imanishi}, {Nakanishi}, {Meier}, {Wada}, {Kawakatu}, \&
  {Nakajima}}]{Izumi2017}
{Izumi}, T., {Kohno}, K., {Fathi}, K., {et~al.} 2017, \apjl, 845, L5

\bibitem[{{Izumi} {et~al.}(2018){Izumi}, {Wada}, {Fukushige}, {Hamamura}, \&
  {Kohno}}]{Izumi2018}
{Izumi}, T., {Wada}, K., {Fukushige}, R., {Hamamura}, S., \& {Kohno}, K. 2018,
  \apj, 867, 48

\bibitem[{{Kalnajs}(1991)}]{Kalnajs1991}
{Kalnajs}, A.~J. 1991, in Dynamics of Disc Galaxies, ed. B.~{Sundelius}, 323

\bibitem[{{Kormendy} \& {Ho}(2013)}]{Kormendy2013}
{Kormendy}, J. \& {Ho}, L.~C. 2013, \araa, 51, 511

\bibitem[{{Maloney} \& {Begelman}(1997)}]{Maloney1997}
{Maloney}, P.~R. \& {Begelman}, M.~C. 1997, \apjl, 491, L43

\bibitem[{{Masters} {et~al.}(2011){Masters}, {Nichol}, {Hoyle}, {Lintott},
  {Bamford}, {Edmondson}, {Fortson}, {Keel}, {Schawinski}, {Smith}, \&
  {Thomas}}]{Masters2011}
{Masters}, K.~L., {Nichol}, R.~C., {Hoyle}, B., {et~al.} 2011, \mnras, 411,
  2026

\bibitem[{{Mezcua}(2017)}]{Mezcua2017}
{Mezcua}, M. 2017, International Journal of Modern Physics D, 26, 1730021

\bibitem[{{Miller} {et~al.}(2015){Miller}, {Kaastra}, {Miller}, {Reynolds},
  {Brown}, {Cenko}, {Drake}, {Gezari}, {Guillochon}, {Gultekin}, {Irwin},
  {Levan}, {Maitra}, {Maksym}, {Mushotzky}, {O'Brien}, {Paerels}, {de Plaa},
  {Ramirez-Ruiz}, {Strohmayer}, \& {Tanvir}}]{Miller2015}
{Miller}, J.~M., {Kaastra}, J.~S., {Miller}, M.~C., {et~al.} 2015, \nat, 526,
  542

\bibitem[{{Miyamoto} {et~al.}(2017){Miyamoto}, {Nakai}, {Seta}, {Salak},
  {Nagai}, \& {Kaneko}}]{Miyamoto2017}
{Miyamoto}, Y., {Nakai}, N., {Seta}, M., {et~al.} 2017, \pasj, 69, 83

\bibitem[{{Mortlock} {et~al.}(2011){Mortlock}, {Warren}, {Venemans}, {Patel},
  {Hewett}, {McMahon}, {Simpson}, {Theuns}, {Gonz{\'a}les-Solares}, {Adamson},
  {Dye}, {Hambly}, {Hirst}, {Irwin}, {Kuiper}, {Lawrence}, \&
  {R{\"o}ttgering}}]{Mortlock2011}
{Mortlock}, D.~J., {Warren}, S.~J., {Venemans}, B.~P., {et~al.} 2011, \nat,
  474, 616

\bibitem[{{Pringle}(1996)}]{Pringle1996}
{Pringle}, J.~E. 1996, \mnras, 281, 357

\bibitem[{{Renaud} {et~al.}(2015){Renaud}, {Bournaud}, {Emsellem}, {Agertz},
  {Athanassoula}, {Combes}, {Elmegreen}, {Kraljic}, {Motte}, \&
  {Teyssier}}]{Renaud2015}
{Renaud}, F., {Bournaud}, F., {Emsellem}, E., {et~al.} 2015, \mnras, 454, 3299

\bibitem[{{Sanders} \& {Huntley}(1976)}]{Sanders1976}
{Sanders}, R.~H. \& {Huntley}, J.~M. 1976, \apj, 209, 53

\bibitem[{{Sassano} {et~al.}(2023){Sassano}, {Capelo}, {Mayer}, {Schneider}, \&
  {Valiante}}]{Sassano2023}
{Sassano}, F., {Capelo}, P.~R., {Mayer}, L., {Schneider}, R., \& {Valiante}, R.
  2023, \mnras, 519, 1837

\bibitem[{{Schawinski} {et~al.}(2010){Schawinski}, {Dowlin}, {Thomas}, {Urry},
  \& {Edmondson}}]{Schawinski2010}
{Schawinski}, K., {Dowlin}, N., {Thomas}, D., {Urry}, C.~M., \& {Edmondson}, E.
  2010, \apjl, 714, L108

\bibitem[{{Smaji{\'c}} {et~al.}(2014){Smaji{\'c}}, {Moser}, {Eckart},
  {Valencia-S.}, {Combes}, {Horrobin}, {Garc{\'\i}a-Burillo},
  {Garc{\'\i}a-Mar{\'\i}n}, {Fischer}, \& {Zuther}}]{Smajic2014}
{Smaji{\'c}}, S., {Moser}, L., {Eckart}, A., {et~al.} 2014, \aap, 567, A119

\bibitem[{{Tristram} {et~al.}(2022){Tristram}, {Impellizzeri}, {Zhang},
  {Villard}, {Henkel}, {Viti}, {Burtscher}, {Combes}, {Garc{\'\i}a-Burillo},
  {Mart{\'\i}n}, {Meisenheimer}, \& {van der Werf}}]{Tristram2022}
{Tristram}, K. R.~W., {Impellizzeri}, C.~M.~V., {Zhang}, Z.-Y., {et~al.} 2022,
  \aap, 664, A142

\bibitem[{{Venemans} {et~al.}(2016){Venemans}, {Walter}, {Zschaechner},
  {Decarli}, {De Rosa}, {Findlay}, {McMahon}, \& {Sutherland}}]{Venemans2016}
{Venemans}, B.~P., {Walter}, F., {Zschaechner}, L., {et~al.} 2016, \apj, 816,
  37

\bibitem[{{Vitral} {et~al.}(2023){Vitral}, {Libralato}, {Kremer}, {Mamon},
  {Bellini}, {Bedin}, \& {Anderson}}]{Vitral2023}
{Vitral}, E., {Libralato}, M., {Kremer}, K., {et~al.} 2023, \mnras, 522, 5740

\bibitem[{{Vitral} \& {Mamon}(2021)}]{Vitral2021}
{Vitral}, E. \& {Mamon}, G.~A. 2021, \aap, 646, A63

\bibitem[{{Yang} {et~al.}(2020){Yang}, {Wang}, {Fan}, {Hennawi}, {Davies},
  {Yue}, {Banados}, {Wu}, {Venemans}, {Barth}, {Bian}, {Boutsia}, {Decarli},
  {Farina}, {Green}, {Jiang}, {Li}, {Mazzucchelli}, \& {Walter}}]{Yang2020}
{Yang}, J., {Wang}, F., {Fan}, X., {et~al.} 2020, \apjl, 897, L14

\end{thebibliography}

%%%%%%%%%%%%%%%%%%%%%%%%%%%%%%%%%%%%%%%%%%
\end{document}